\documentclass[pre,twocolumn]{revtex4}
\usepackage{graphicx,t1enc}
\usepackage{amssymb,amsmath}
\usepackage{verbatim}

\begin{document}

 \title{Cipolla' s game: playing under the laws of human stupidity}

\author{Joel Kuperman} 
 \affiliation{Facultad de Matem\'atica, Astronom\'{\i}a y F\'{\i}sica, Universidad Nacional de C\'ordoba, Ciudad Universitaria, 5000 C\'ordoba, Argentina}

\author{Donny R. B\'arcenas}
\affiliation{Instituto Balseiro, R8400AGP Bariloche, Argentina.}

\author{Marcelo N. Kuperman}
\affiliation{Consejo Nacional de Investigaciones Cient\'{\i}ficas y T\'ecnicas \\
Centro At\'omico Bariloche (CNEA) and Instituto Balseiro, R8400AGP Bariloche, Argentina.}

\begin{abstract}

In this work we present an evolutionary game inspired by the work of Carlo Cipolla entitled \textit{The basic laws of human stupidity}. The game 
expands the classical scheme of two archetypical strategies, collaborators and defectors, by including two additional strategies. One of 
these strategies is associated to a stupid player that according to Cipolla is the most dangerous one as it undermines the global wealth of the 
population.
By considering a spatial evolutionary game and imitation dynamics that go beyond the paradigm of a rational player we explore the impact of Cipolla' 
s ideas and analyze the extent of the damage that the stupid players inflict on  the population.
\end{abstract}

\date{\today}

\maketitle

\section{Introduction}

When around  1976 Cipolla formulated the fundamental laws of human stupidity, he  was
being sarcastic and trying to build a cartoonish image of the human society. However, his ideas contained some aspects that constituted an adjusted 
characterization of the type of behaviors displayed in interpersonal relationships.
In his work, published in 1988 \cite{cipolla88} Cipolla points at describing the personal interactions in terms of benefits and damages derived from 
any transaction, conceptually going beyond monetary aspects exclusively. He pointed at the concept of stupidity as seen within a social 
context, and to establish a proper frame for his ideas he classified the behavior that an individual may display within a social context into four 
groups. These groups are the intelligent (I), the bandit (B), the unsuspecting (U) and the stupid (S).
The difference between them arises from the inclination to produce benefits or harms for oneself and for others in any interaction.

\begin{figure}%[!h]
\includegraphics[clip=true, width=8cm]{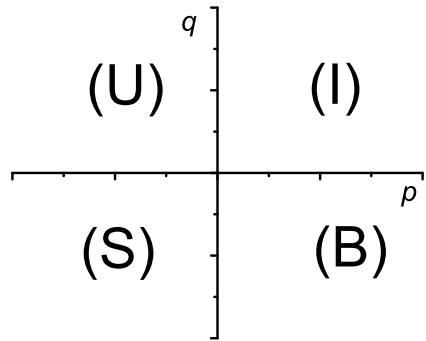} \caption{
Location of each strategy in the \textit{(p,q)} space.}
\label{esque1}
\end{figure}

It should be pointed out that in Cipolla' s work the concepts of stupidity and intelligence are lax and do not intend to refer to any cognitive 
abilities of the subjects. Group (I) consists of individuals who when they interact with others produce a mutual benefit.
Group (B) is composed by selfish individuals who seek individual benefits without hesitating to cause harm to others.
Group (U) represents a type of altruistic individual who seeks the wealth of others even at the expense of self-inflicted harm. Finally,  the (S) 
group contains the individuals that not only cause harm to others but also to themselves.
In order to mathematically represent the behaviors associated to each group, it is possible to choose two parameters: the 
gains or losses that an individual causes to him or herself, $p$, and the gains or  losses that an individual inflicts on others, $q$ .
These four groups are then defined by the range of values adopted  by  $p$ and $q$ as follows :
\begin{itemize}
\item[S] : $p_s \leq 0$ y $q_s < 0$
\item[U]: $p_u \leq 0$ y $q_u\geq 0$
\item[I]: $p_i > 0$ y $q_e \geq 0$
\item[B]: $p_b > 0$ y $q_b < 0$
\end{itemize}
Figure \ref{esque1} shows the location of each strategy on the $(p,q)$ plane.
Besides the previous classification of the population into four groups, the central point in Cipolla' s work is the enunciation of \textit{The Basic 
Laws of Human Stupidity}, 
listed bellow and quoted from \cite{cipolla88}

\begin{enumerate}

\item \textit{Always and inevitably everyone underestimates the number of stupid individuals in circulation.}
\item \textit{The probability that a certain person be stupid is independent of any other characteristic of that person.}
\item \textit{A stupid person is a person who causes losses to another person or to a group of persons while himself deriving no gain and even 
possibly 
incurring losses.}
\item  \textit{Non-stupid people always underestimate the damaging power of stupid individuals. In particular non-stupid people constantly forget that 
at all times and places and under any circumstances to deal and/or associate with stupid people always turns out to be a costly mistake.}
\item    \textit{ A stupid person is the most dangerous type of person.}
\end{enumerate}
\textbf{Corollary:} \textit{a stupid person is more dangerous than a pillager.}

\vspace{.3cm}
The mathematical characterization of the four groups together with the fundamental laws, inspire us to formulate an evolutionary game, that we call 
The Cipolla's game. Each of the four groups described above is associated to a possible strategy and the corresponding payoff matrix is built in 
terms 
of the outcome of the interactions between them. The values of this matrix are loaded in the following table, that indicates which is the payoff of 
the strategy at the file when competing with the strategy at the column
\begin{table}[h]
\centering
\begin{tabular}{c| c| c|c |c}
 &{\bf S} & {\bf U} &{\bf I} & {\bf B}\\
%heading
\hline
{\bf S} & $p_s +q_s$ & $p_s+q_u$& $p_s+q_i$ & $p_s+q_b$  \\
\hline
{\bf U} & $p_u +q_s$ & $p_u +q_u$& $p_u +q_i$ & $p_u +q_b$  \\
\hline
{\bf I} & $p_i +q_s$ & $p_i +q_u$& $p_i +q_i$ & $p_i +q_b$  \\
\hline
{\bf B} & $p_b +q_s$ & $p_b +q_u$ & $p_b +q_i$ & $p_b +q_b$
\end{tabular}
\caption{Payoff Table} \label{tabla1}
\end{table}

Once the strategies and the payoffs are defined, we propose an evolutionary game, which dynamics can be associated with 
that of the replicator. In the following we will consider that the (B) group is the one that gets the highest self reward $p$ and thus being a bandit 
has certain incentives. As we will show later, the resulting game has a unique strict Nash equilibrium, the strategy (B). If we consider a mean field 
model described by the usual replicator equations,  (B) is the only trivial stable steady state and thus the population converges to an homogeneous 
group of bandits. 

In order to have a richer dynamics we can consider the sub game in which only the strategies  (I) and (B) participate and choose 
the values of the payoff matrix in order to get a Prisoner's Dilemma (PD) . While this does not add anything to the previous observation regarding 
the Nash equilibrium when formulating  mean field 
equations, previous works have shown that when considering an underlying network defining the topology of the interaction between players, the 
results can change. It has been observed that a departure from the assumption of a well mixed population promotes the emergence of cooperation in the 
classical PD game, at least for certain network topologies and a range of values for the payoffs of the competing strategies \cite{dur,kup3}.
Based on these results, one of the objectives of this work is to understand how the  topology affects the dynamics of the game. 
For that we introduce a spatially extended game and  consider that the topology of the interactions between players is described by a 
network. In such a case  each player plays with its neighbors and the  decision to update its strategy is based only on the local information 
collected throughout the  neighborhood. There is a plethora of network topologies from which we can choose the substrate. In this work  we will 
focus on a family of networks  that are likely to enhance the effects on the propagation of a cooperative behaviour such as (I) due to the local 
character of the dynamics.  These networks. described in \cite{watts98} and \cite {kup2} present a topology that  varies according 
to the value of the disorder parameter. In particular, there are two quantities of interest such as the clustering coefficient and the average path 
length though we will focus on the first. 

On the other hand, we must define an imitation dynamics associated to the evolution of the strategies distribution among the population. 
The simplest assumption is to think of a deterministic imitation. In each round a given player, the focal one, plays with all its 
neighbors, while each of its neighbors does the same with  their own. After that round the focal player analyses its performance or earnings and 
compares them with that of its neighbors. Then, it adopts the strategy of the player with the highest gain. In case of tie the choice  is decided at 
random. This update dynamics is the simplest one, representing a deterministic imitation and closely linked to the replicator dynamics \cite{hofb}. 
Adding non deterministic aspects can lead to  more interesting dynamics, but will also screen the topological effects.

Either case, deterministic or not,  is not considering the nature of the players. The dynamics originally proposed is based on the idea 
of a rational player, who seeks its own  benefit above all. This is directly associated with 
the characteristics of a (B) player but not with the rest. For example, if we attain to the laws of Cipolla, a player (S) will not be 
interested in earning a higher profit and could  ignore what happens with its neighborhood, that is, it  could  stay immutable and not change the 
strategy at all or even imitate the strategy of the neighbor that has caused the higher loss to the rest.

One way to include this in the imitation dynamics is to consider different inclinations no to behave as dictated by rationality according to the 
nature of the groups. In the following sections we discuss this possibility.

\section{Mean field results}

In this section we analyze the replicator dynamics, under the assumption  of a well mixed population. 
First we introduce the payoff matrix
\[A=\begin{pmatrix}
p_s +q_s & p_s+q_u& p_s+q_i & p_s+q_b\\
p_u +q_s & p_u +q_u& p_u +q_i & p_u +q_b \\
p_i +q_s & p_i +q_u& p_i +q_i & p_i +q_b\\
p_b +q_s & p_b +q_u & p_b +q_i & p_b +q_b
\end{pmatrix}\]

As stated in the introduction, we  choose the values of the payoff matrix so that the sub game ($ I, B $) is a Prisoner's dilemma. 
In that case we need $$ p_b + q_i> p_i + q_i> p_b + q_b> p_i + q_b. $$ Given that $ q_b <0 $ and $q_i >0 $ it is enough to choose $ p_b> p_i $.

The equations for the evolution of the density of each strategy $x_k$  are 
\begin{equation}
 \dot x_k= x_k ([A\vec x]_k-A\vec x A)
 \label{repli}
\end{equation}
where $[A\vec x]_k=\sum_j a_{kj} x_j$ and $a_{kj}$ are the elements of $A$.
From now on we associate the subindices 1,2,3,4 with $s$,$u$,$i$,$b$ respectively.

We can simplify the calculations by making use of one property of the replicator equations that says that the addition of a constant $c_k$ to the 
$k$-th column of $A$ does not change Eq (\ref{repli}) (when restricted to the simplex where the relevant dynamics occurs) \cite{hofb}. We can use 
then 
\[ B= \begin{pmatrix}
p_s & p_s& p_s & p_s\\
p_u & p_u& p_u  & p_u \\
p_i & p_i& p_i & p_i \\
p_b & p_b & p_b & p_b
\end{pmatrix}\]
and show that the dynamics is solely defined by the $p_k$ values.  Eq. (\ref{repli}) can now be written in a much simpler form
\begin{equation}
 \dot x_k= x_k (p_k-\sum_j x_j p_j )
 \label{repli2}
\end{equation}

This system has four relevant steady solutions corresponding to the survival of a single 
strategy. The Jacobian of the system is

 \begin{widetext}\[\begin{pmatrix}
(1-x_s)p_s-\bar p& -x_s p_u& -x_s p_i & -x_s p_b\\
-x_d p_s & (1-x_d) p_u-\bar p& -x_u p_i & -x_u p_b \\
-x_i p_s & -x_i \pi_d& (1-x_i)p_i-\bar p& -x_i p_b \\
-x_b p_s & -x_b \pi_d& -x_b p_i &(1-x_b)p_b-\bar p 
\end{pmatrix} \] \end{widetext} 
with $\bar p=\sum_j x_j p_j$. Considering that the steady states correspond to only one of the $x_k$ being equal to 1 and the rest equal to 0, the 
eigenvalues for a state when $x_k=1$ and $x_j=0$ for $j \neq k$ are 
$$(1-\delta_{k,j}) p_j-p_k.$$
It is straightforward to conclude  that the only stable steady state, when $B$ has four negative eigenvalues,  is the one corresponding to the 
survival 
of the strategy with the highest $p_k$. Thus, when considering a mean field model, the population converges to an homogeneous  group  of bandits.

\section{Dynamics on networks}

During the last decade many authors began studying evolutionary spatial games to overcome the limitations associated with the assumption that 
players were always part of a well-mixed population \cite{now0, now3, sza2}. These works showed that the
evolutionary behaviour and survival of the populations of each strategy  might be affected by the underlying topology of links between players \cite 
{kup2, kup3,now1,kup1,dur, roc}.

The fact that strategies not associated with the Nash equilibrium can survive by forming clusters and gain certain advantage from this has been 
analyzed in several works where the classical cooperative (C) and non-cooperative (D) strategies are considered \cite {now3, roc, oht, doe, 
hauert,con,ass, lei, ift, san2, vuk2, gan}.

We can gain some intuition about what is happening by the following reasoning. If (C) nodes can exploit 
the advantages of mutual cooperation, the effect of clustering would be to protect 
the internal (C) nodes  from the presence of the (D) nodes at the border. Since (D) can only get advantage from its interaction with the (C), only 
those defectors located on the border of a group of cooperators can have benefits, while the  grouped (C) obtain benefits from the mutual cooperation.

If the (C) nodes  at the boundaries of the cluster notice that the  cooperators inside  do better than the (D) outside they will not be 
tempted  to change their strategy and  they might even succeed to expand the cooperative strategy towards the defective population. However, this 
phenomenon is strongly dependent  on the relative values of the payoff of (C) and (D) when playing against (C)  and on the structure of the network. 
The most relevant feature in this regard is the clustering coefficient, which 
measures the mean connectiveness between the members of a node' s neighborhood. Ultimately, it is the existence of local transitive relationships, 
closely related to clustering \cite{was}, what defines the possibility of survival and expansion of small cooperator groups \cite{kup3}.

In this work we consider regular networks with a tunable degree of  disorder that translates into different values ​​of clustering and path
length. By construction, these networks are regular because all the nodes have the same number of neighbors. To build them we  use a modified 
algorithm based on the one originally proposed in \cite {watts98} that  maintains the regularity \cite{kup2}.

\begin{figure}%[!h]
\includegraphics[clip=true, width=7cm]{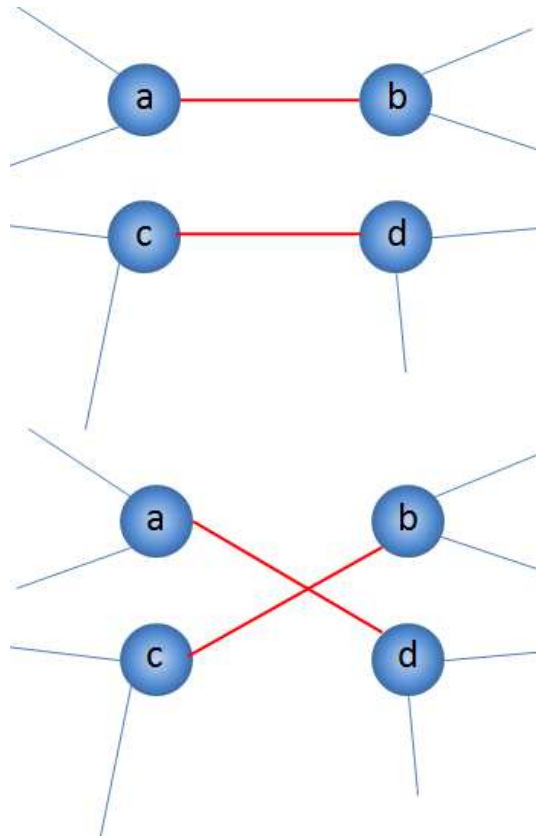} \caption{
Algorithm of construction of the k-SW networks. In this example, a single change is depicted.  All nodes have degree equal to 3. Initially node 
\textbf{a} and \textbf{b}; and nodes \textbf{c} and \textbf{d} are connected. 
After the exchange, \textbf{a} is connected to \textbf{d} and \textbf{c} is connected to \textbf{b}. The degree of the nodes has not changed.}
\label{net}
\end{figure}

The usual algorithm of construction of WS networks is as follows: starting from an ordered network, each link is rewired  with a certain fixed 
probability, preserving one of its adjacent node but connected  the other extreme to  a random one. Double and self links are not allowed. Though
the algorithm conserves the total number of links, at the end of the process the degree of each node is statistically
characterized by a binomial distribution. As we are interested in filtering any effect related to changes in
the size of the neighborhoods we modify the original WS
algorithm to constrain the resulting networks to a subfamily
with a delta shaped degree distribution. We call this
family of networks the k-Small World (k-SW) networks,
where 2k indicates the degree of the nodes. The procedure is schematized in Fig \ref{net},

The construction procedure begins again with a regular ordered network which structure is broken by
a sequential exchange of the nodes attached to the ends of two randomly chosen links.
Starting, for example, from an ordered ring network, each link is subject to the possibility of exchanging one of its
adjacent nodes with another randomly chosen link with probability $ \pi_d $. Thus, to proceed with the reconnection of the network
we choose two couples of linked nodes (or partners) rather than one. If  we accept  to switch the partners  we get  two new pairs of 
coupled nodes. In this way all the nodes preserve their degree
while the process of reconnection ensures the introduction
of a certain degree of disorder.

\subsection{Simple deterministic dynamics}

We consider first the simplest dynamics. A chosen  player plays with it neighbors, who in turn also play with the members of their neighborhoods. 
After that round, the chosen player imitates the most successful neighbor.  But at this point we introduce a slight variation. While the imitation of 
the most successful will be always the rule for  (B), (I), and 
(U), we will analyze  two different behaviors for (S): one in which it imitates the best neighbor as the other strategies and one in which it never 
changes the strategy. In this case  an (S) player remains always as (S). We will call the first case  \textit{no-frozen} and the second one 
\textit{frozen}. 

As shown in \cite{kup3, dur}, we need to take into account that in order for the game to have a non trivial dynamics
and allow the survival of strategies other than the Nash equilibrium such as (I), the quotient $\frac{p_m}{p_i}$ must not exceed a certain 
threshold value that depends on the topology of the underlying network, especially on the clustering of the nodes and the mean degree.
Thus, we will fix the values of all the parameters letting $p_m$ vary within a proper range.

The chosen values are
\begin{table}[h]
\centering
\begin{tabular}{| c| c|c |c|c|c|c|c|}
\hline
 $x_i$ &$x_b$ &$x_d$ &$x_e$&$y_i$&$y_b$&$y_d$&$y_e$\\
\hline
 1 &[1.1,2] &[-2,-1] &[-2,-1]&1&-1&1&-1\\
 \hline
\end{tabular}
\caption{Chosen values for the payoff matrix} \label{tabla2}
\end{table}

The main goal of this work is to characterize the influence of the (S) strategy on the dynamics of the strategy profile of the  population. 
This is the main rationale to compare the results derived from the \textit{frozen} and \textit{no frozen} dynamics. 
Considering the first two laws it would be interesting to analyze the effect of the proportion of (S) players among the population. Therefore we 
also  take different initial fractions of (S) and analyze the effect they may have on the global wealth of the population.

Here we show results corresponding to networks with  $10^5$ nodes and degree 8, although we have tested different degrees to 
ensure that this choice does not affect the generality of the results. The only constraint is that the network should be diluted, i.e. 
a relatively low mean  degree.

In all the cases we have verified the convergence to a global steady state, with sometimes  negligible local dynamics. Once this steady state is 
reached, we measure the fraction of individuals in each strategy, $x_k$. We will show that despite the Nash equilibrium  of the game 
is the pure strategy (B), the spatial effects can make the (I) strategy survive. In most cases, except when the fraction of (S) is maintained fixed, 
the populations of (S) and (U)  disappear.

To analyze the effect of the network topology on the final state we consider several values of $\pi_d$ and to understand the role of (S), we start 
with different  fractions of its population. 

To characterize the steady state we measure the ratio  $x_ i/x_b$ and the total profit that is being generated in the 
population due to the interactions, $<\epsilon>$. When the strategies (U) and (S) are absent in the steady state  both quantities will 
display exactly the same behavior but when at least one of these two strategies survives we will need both to fully recover the information of what 
is happening in the system. 

Across the numerical calculations we verified that the system quickly reaches the steady state after 10000 time steps, each one consisting in $N$ 
rounds of a game between a randomly chosen node and its neighbors.  First we point to analyze the effect of the initial population of (S) players, 
$\rho_s(0)$ and the topology of the network. For this reason we consider several values of  $\rho_s(0)$ and $\pi_d$.
At the beginning of the dynamics,  the fraction  of the rest of the strategies is the same, $(1-\rho_s(0))/3$.
We have scanned the results for several values of the parameters $p_k$ and $q_k$, and found two distinct situations.
If we take $1.1<p_b< 2.0$ the (I) strategy can always survive thanks to the advantage it can get from the formation of clusters of (I) individuals 
that collaborate with each other, giving them advantages over (B). When $p_b>2$ this advantage disappears and the population of (I) tends to 0. 
The (S) strategy, when present, does not have this advantage and  disappears, just like (U), unless we consider the \textit{frozen} dynamics.

\begin{figure}%[!h]
\includegraphics[clip=true, width=9cm]{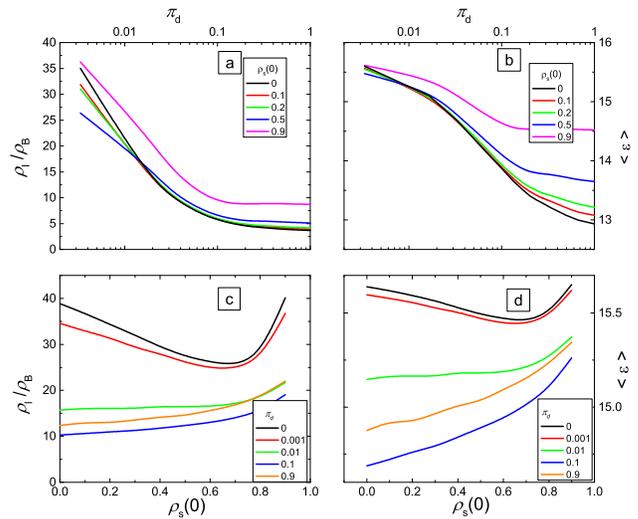} \caption{
These plots display the results for  the \textit{no frozen} dynamics. In the top plots we show  the results as a function of $\pi_d$ and 
different values of $\rho_s(0)$.  a) This plot shows the ratio $\rho_i / \rho_b$ at the steady state. b) This figure shows the mean 
gain at the steady state.  The bottom plots display the  results as a function of $\rho_s(0)$ for different values of $\pi_d$. c) This plot shows the 
ratio $\rho_i /\rho_b$ at the steady state. d) This figure shows the mean gain at the steady state. In these plots the thick line correspond to 
$\rho_s(0)=0$. $p_b=1.2$}
\label{cip3}
\end{figure}

First, we study the \textit{no frozen} dynamics. 
Figures \ref{cip3}.a and \ref{cip3}.b show the values adopted by the ratio $\rho_i / \rho_b$ and the mean gain of the population $<\epsilon>$ at the 
steady state respectively as a function of $\pi_d$. We find  that effectively the (S) and (U) fractions fall to zero and the steady state shows a 
weak dependence of the initial fraction of (S). The game ends up being a prisoner's dilemma and the results qualitatively agree with those obtained 
in other works for this case. The fraction of (I) decreases as $\pi_d$ increases \cite{kup3}. However, the initial fraction of (S) affects the final 
state in a non trivial way. Except for the lowest values of $\pi_d$, it seems to have an effect contrary to the one predicted by Cipolla, as the 
increase of  $\rho_s(0)$ leads to a steady state with a higher ratio of cooperators and even a higher main global gain. We will propose later an 
explanation for this effect. This can be more clearly observed in Figures \ref{cip3}.c and \ref{cip3}.d, where 
we show the values adopted by $\rho_i / \rho_b$ and $<\epsilon>$ as a function of $\rho_s(0)$. The crossover observed in figs. \ref{cip3}.a and 
\ref{cip3}.b is reflected in the change of slope of the curves according to the values of $\pi_d$. 

In this analysis we also include the case when $\rho_s(0)=0$, that helps us to evaluate the effect of $\rho_s(0)\neq 0$. We observe that for the 
lowest values of $\pi_d$ and $\rho_s(0)$ the population is harmed by the presence of (S). This scenario seems to change for higher values of $\pi_d$ 
or when $\rho_s(0)$ is high enough. As mentioned before, we will provide an explanation after studying the \textit{frozen} case.

In the former example, the populations of (S) and (U) decay to reach extinction. 

Next, we may think of an  alternative imitation dynamics  that 
might seem to be a closest interpretation of Cipolla' s laws. We  now consider that the population of (S) does not change its 
strategy throughout the evolution of the strategies of the the rest of the population. Note that  the unlikely adoption of the strategy (S) is not 
forbidden. 

\begin{figure}%[!h]
\includegraphics[clip=true, width=9cm]{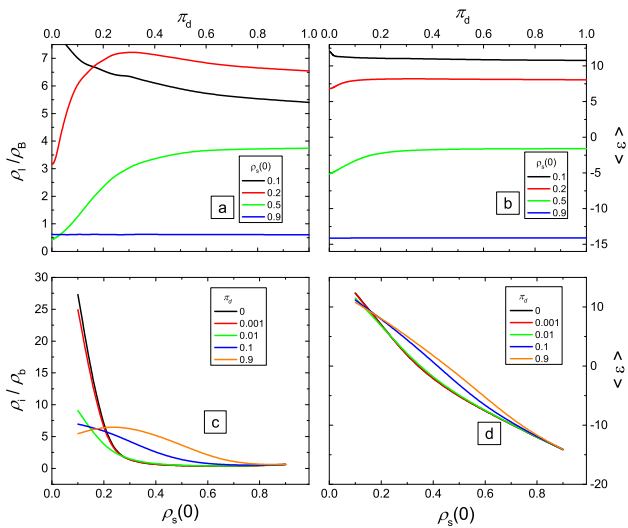} \caption{
These plots display the results for  the \textit{frozen} dynamics. In the top plots we show  the results as a function of $\pi_d$ and 
different values of $\rho_s(0)$.  a) This plot shows the ratio $\rho_i / \rho_b$ at the steady state. b) This figure shows the mean 
gain at the steady state.  The bottom plots display the  results as a function of $\rho_s(0)$ for different values of $\pi_d$. c) This plot shows the 
ratio $\rho_i /\rho_b$ at the steady state. d) This figure shows the mean gain at the steady state.  $p_b=1.2$}
\label{cip1}
\end{figure}

The results are shown in Figures \ref{cip1}, with a correspondence between the panels of Figures \ref{cip3} and this one. 
We see that in most cases the value of $ \pi_d $ increases, the final fraction of (I), $\rho_i$, increases too. This is not the case for the lowest 
values of $\rho_s(0)$, when the results are similar to what has been observed for $\rho_s(0)=0$.

These results give us a hint of what could be happening that could explain why in the \textit{no frozen} case, the highest initial fraction of (S) 
favors the survival of (I). When confronted with an (S) player, the (I) will never change its strategy. The only temptation for a change comes from a 
possible higher payoff only attainable by a (B) player. Thus, the (S) population is screening or isolating the (I) players, letting them to clusterize 
and eventually propagate their strategy. In the \textit{no frozen} case, this transient phenomenon  leads a an increase in the ratio $\rho_i / 
\rho_b$. In the \textit{frozen} case this effect is limited by the permanent presence of (S), that partially inhibits the propagation of both 
strategies. 

But in the presence of (S) player in the steady state, the ratio $\rho_i / \rho_b$ is not giving us a proper information of the state of the 
population, as potentially (B) players are being replaced by (S). Thus we analyze the values of $<\epsilon>$. We observe 
that unlike in the \textit{no frozen} case the greater the initial fraction of (S), the worse  is the performance of the population. And also, we can 
see that when  the mean profit is always lower than in the \textit{no frozen} case. 
Thus the survival of the (S) population results in a clear global damage.

Some of the results shown in Fig. \ref{cip1} could be explained just by the fact that we are starting with a higher initial  number of individual 
within the frozen (S) population, leaving us with a trivial effect. Given that the population of (S) is maintained frozen, it is not surprising that 
the wealth  of the population decreases with the initial fraction of (S), but the curves displayed in Fig. \ref{cip1}.d show a non linear 
dependence, evidencing  non trivial effects.

As stated in  previous works \cite{kup3, dur}, the possibility of survival of (I) depends on the ratio between the payoffs received by 
strategies (I) and (B) when confronting another (I), that is $ (p_b + q_i)/( p_i + q_i) $. As this ratio grows, the surviving fraction of (I) 
population decreases. For both cases we verified that for $p_b\ge 2$,  only (B) players survive, except for the frozen population of (S) in the 
corresponding case.

We note that in all the cases studied above, the population of (U) disappears.

\subsection{Specific dynamics}

In the previous section we considered a differentiated imitation dynamics only for the (S) strategy. Here we explore an expansion of this idea by 
considering a specific imitation dynamics for each strategy, always inspired by the principles that characterize each of them.

\begin{figure}%[!h]
\includegraphics[clip=true, width=9cm]{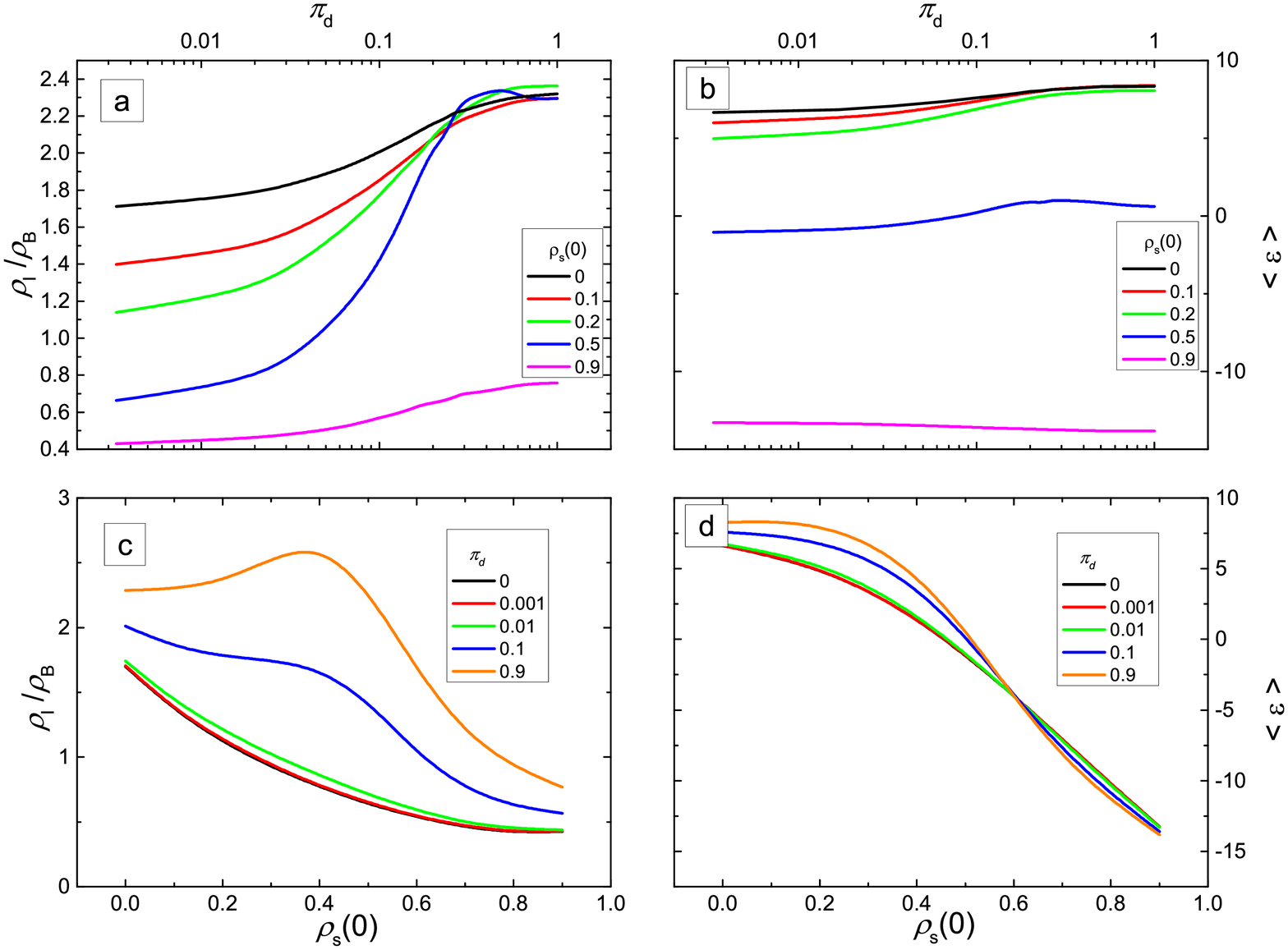} \caption{These plots display the results for  the \textit{non frozen} dynamics and 
specific imitation behavior. In 
the top plots we show  the results as a function of $\pi_d$ and 
different values of $\rho_s(0)$ when the imitation dynamics is differentiated.  a) This plot shows the ratio $\rho_i / \rho_b$ at the steady state. 
b) This figure shows the mean 
gain at the steady state.  The bottom plots display the  results as a function of $\rho_s(0)$ for different values of $\pi_d$. c) This plot shows the 
ratio $\rho_i /\rho_b$ at the steady state. d) This figure shows the mean gain at the steady state. $p_b=1.2$}
\label{cip4}
\end{figure}

Among the four groups defined by Cipolla  only (B) behaves like a rational player, always looking for the individual wealth above all and 
therefore always  imitating the neighbor with the highest profit. On the opposite side, the  $ U $ group presents an 
altruistic nature, seeking for the benefit of the other. In that sense, we may assume that such player will try to  imitate the neighbor who 
generates the greatest profit for the the rest, irrespectively of the own profit associated to that change.

In the previous section we consider two possibilities for the imitation behavior of (S): it  could o could not change its behaviour. In the present 
case we will also consider these two options but in case it changes its strategy, it will not act as a rational player. We assume that the need  
to generate damage, regardless of the costs, is rooted in its  nature. Following this premise it will imitate the neighbor that produces  the 
greatest loss or minor gain in its  neighborhood.

\begin{figure}%[!h]
\includegraphics[clip=true, width=9cm]{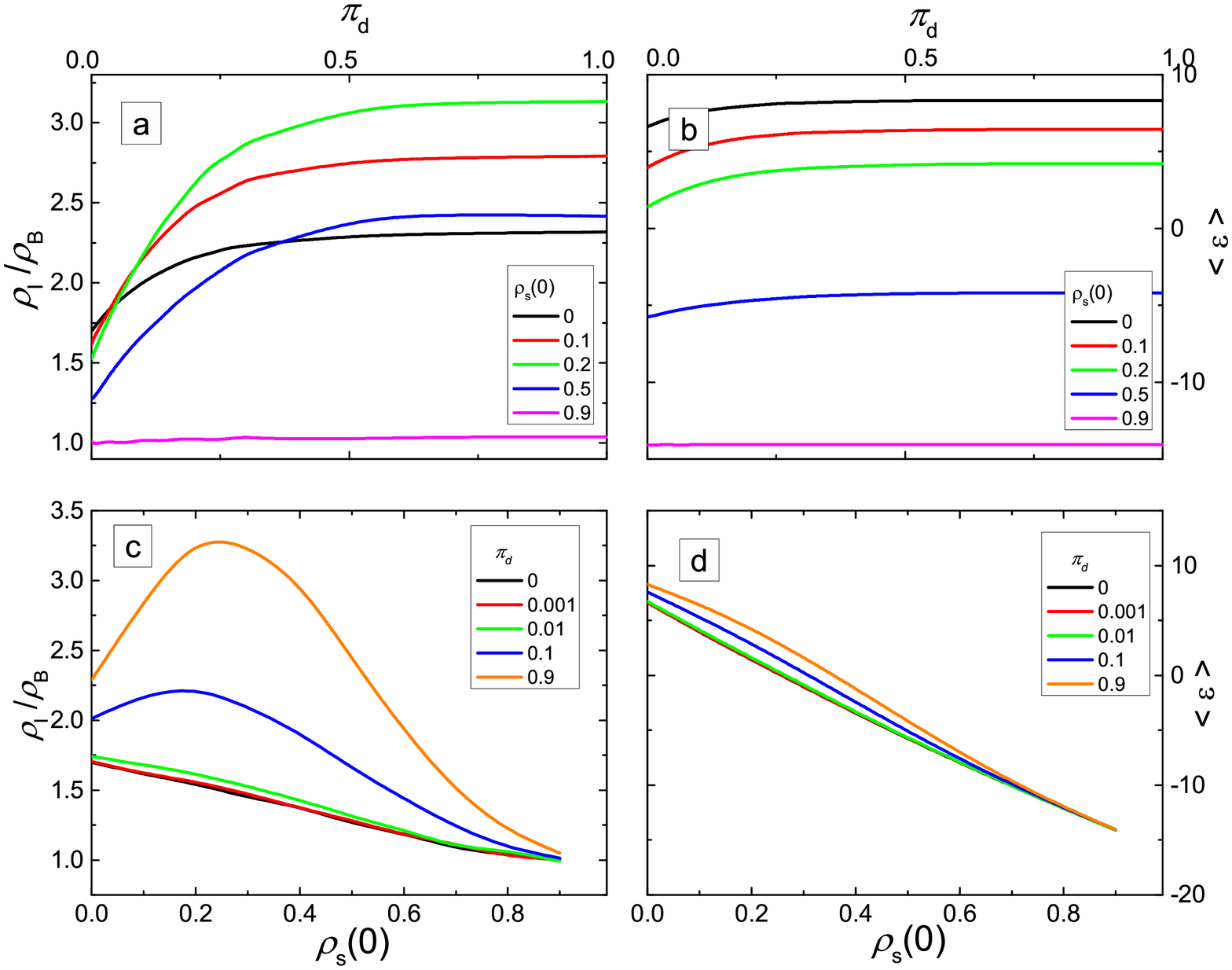} \caption{These plots display the results for  the \textit{frozen} dynamics and 
specific imitation behavior. In 
the top we plots show  the results as a function of $\pi_d$ and 
different values of $\rho_s(0)$ when the imitation dynamics is differentiated.  a) This plot shows the ratio $\rho_i / \rho_b$ at the steady state. 
b) This figure shows the mean 
gain at the steady state.  The bottom plots display the  results as a function of $\rho_s(0)$ for different values of $\pi_d$. c) This plot shows the 
ratio $\rho_i /\rho_b$ at the steady state. d) This figure shows the mean gain at the steady state. $p_b=1.2$}
\label{cip5}
\end{figure}

Finally, we consider that the (I) group shows some traces of altruism but not a the cost of self generating a loss.  So it  will seek not to suffer a 
loss but at the same time to be involved with the generation of a global profit. So it will imitate the neighbor who generates the greatest global 
profit and at the same time does not involve its own loss.

So, as in the previous section, we will have a \textit{no frozen} and a \textit{frozen} case. As will be shown, the results for both cases present a 
new feature, the survival of the (U) population. 

Both cases show results  qualitatively very similar to what we have obtained for the \textit{frozen} dynamics in the previous example, reflecting 
that the dynamics chosen for the (S) groups ensures its survival.

Figures \ref{cip4} show the results for the \textit{no frozen} dynamics. The new imitation behavior adopted by (S) prevents it from changing the 
strategy, indicating that even at a local scale, the (S) player is the one causing the greater loss. 
Despite the similarities, the mean profit of the population is always higher for the \textit{no frozen} case, mainly due to the fact that the 
presence of (I) players is higher, as can be observed in Figs. \ref{cip5}. Also, in the \textit{no frozen} case there is a  decrease of the (S) 
population, reaching steady fractions verifying  $\rho_s \approx \rho_s(0)^2$.

\begin{figure}%[!h]
\centerline{\includegraphics[clip=true, width=7cm]{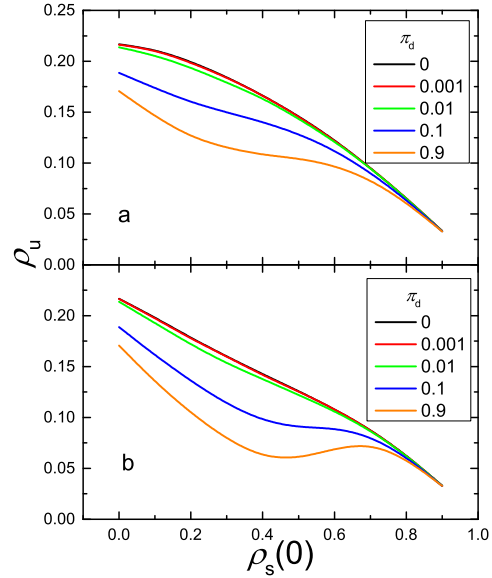}} \caption{Steady fraction of (U), $\rho_u$ as a function of $\rho_s(0)$. for the 
\textit{frozen } (a) and \textit{no frozen} (b) cases}
\label{propu}
\end{figure}

The main difference between the former results and the new ones resides in the fact that now, a small 
population of (U) can persist. This is shown if Fig \ref{propu} where the steady fraction 
of (U), $\rho_u$ is depicted. The figures show the $frozen$ and $no frozen$ cases, for several values of $\pi_d$ and as a function of 
$\rho_s(0)$.

\section{Conclusions}

In this work we present a mathematical interpretation and analysis of the ideas introduced by Carlo Cipolla in \cite{cipolla88}.
The adopted formalism is based on the formulation of an evolutionary game which payoff matrix is a direct translation of the definition of the four 
groups characterizing the nature of human transactions. We have shown that the resulting game has a unique Nash equilibrium and thus the evolution 
of the strategies under the replicator dynamics leads to a trivial solution corresponding to an homogeneous population of bandits.
Based on previous results on spatial cooperative games, we adopted payoff values that let us identify some features of the present game with a 
Prisoner' s Dilemma. In addition to this, we explored a spatial version of the game, by considering a selected family of underlying regular networks.
These networks are characterized by a single disorder parameter and the degree of the nodes. 

The analysis of the spatial version of the game presented interesting results that let us reveal the mathematical structure behind  the ideas of 
Cipolla. 

According to Cipolla' s laws, the number of stupids cannot be estimated. In order to explore the possibility of a critical fraction of (S) 
individuals can affect the population, we have explored a range of values in the interval $[0,1]$. We have found that even the smallest fraction of 
stupids produces  a notable effect. Letting aside some subtleties to be explained later, the overall conclusion is 
that as the fifth law establishes, a stupid person is the most dangerous one, even more dangerous than a bandit. This is reflected in the fact that 
in most of  the cases, a higher fraction of (S) lead to a lower global gain, independently of whether the (S) group can or can not change its 
strategy. We found some exceptions where the (S) group seems to play  contradictory effects favoring the propagation of (I) players and leading to  
a higher mean profit. Before explaining this effect we want to address other results that deserve a closer look,  related to the behaviour of 
the ratio between the (I) and (B) group, and the survival of (U) individuals.
We have found that when the (S) players  survive,  their steady fraction  only depends on $\rho_s(0)$, the topology of the 
networks seems to play no role. While this may sound obvious for the frozen dynamics, it is not for the \textit{no frozen} one. However, the topology 
of the network is extremely relevant in defining how the initial (S) population will affect evolution and organization of the final state. The (S) 
initial population together with the topology of the network is what governs the final ratio between the (I) and (B) population, and thus the overall 
gain of the population. In all the cases, the permanent presence of the (S) group undermines the wealth of the population and only a transient 
survival can lead to an overall gain.
This phenomenon is the result of a screening effect played by the (S) population, as they isolate the (I) players from the (B) ones avoiding the 
tempting change from (I) to (B). At the same time, during the transient presence of (S), the (I) group strengthens and may start to propagate towards 
the (B) population. At this point, the (S) populations starts to play the opposite role, as it prevents the (I) group to advance over the (B) 
population. This effects is responsible for the non monotonic shape of the curves observed in Figs. \ref{cip1}.c, \ref{cip4}.c, and \ref{cip5}.c.

In this work we have excluded the possibility that $p_i>p_b$. If such was the case, the structure of the game would be different, leading to a 
trivial homogeneous population of (I) individuals, even in an extended game. We wanted to explore a situation in which there is a social dilemma and 
there is a temptation not to adopt a cooperative strategy, such as the Prisoner's dilemma.

In summary, our works explores the ideas of Cipolla showing that their implementation as a game may lead to interesting and non trivial conclusions, 
in agreement with the proposed laws.

In this work we have only considered deterministic dynamics. The introduction of some 
stochasticity, not only in the imitation dynamics but also in 
the possibility of a spontaneous change of strategy of some players will be analyzed in a future work.


\begin{thebibliography}{00}
 

\bibitem{cipolla88} C. Cipolla,\textit{ Allegro ma non troppo} (1988,Il Mulino, Bologna).

\bibitem{kup3} M.N. Kuperman and S. Risau-Gusman,
%\textit{Relationship between clustering coefficient and the success of cooperation in networks.} 
Physical Review E \textbf{86}, 016104 (2012)


\bibitem{dur} O. Dur\'{a}n  and R. Mulet, 
%\textit{Evolutionary prisoner's dilemma in random graphs.} 
Physica D {\bf 208}, 257-265 ( 2005).

 
\bibitem{watts98}  D.J. Watts and S.H. Strogatz, 
%\textit{Collective dynamics of'small-world' networks.}
Nature {\bf 393}  440-442  (1998).

\bibitem{kup2} M.N. Kuperman and S. Risau Gusman,
%{\em The effect of the topology on the spatial ultimatum game.}
Eur. Phys. Jour. B {\bf 62},
233-238 (2008)

%\bibitem{mella} P. Mella \textit{Intelligence and Stupidity.} Creative Education \textbf{8}, 2515-2534 (2017).

\bibitem{hofb} J. Hofbauer and K. Sigmund \textit{Evolutionary games and population dynamics}.(1998, Cambridge U.P., Cambridge).

\bibitem{now0}  M. A. Nowak and R. M. May,
 %{\em Evolutionary games and spatial chaos.}
 Nature {\bf 359},
826-829 (1992).

\bibitem{now3} M. A. Nowak and R. M. May,
%{\em The spatial dilemmas of evolution.}
Int. J. Bifurcation Chaos, {\bf 3}, 35-78 (1993).


\bibitem{sza2} G. Szab\'{o}  and G. F\'{a}th,
%{\em Evolutionary games on graphs.}
Phys. Rep. {\bf 446}, 97-216 (2006)


\bibitem{now1} M. A. Nowak,
%{\em Five rules for the evolution of cooperation.}
Science {\bf 314}, 1560-1563 (2006).


\bibitem{kup1} G. Abramson and M. N. Kuperman,
% {\em Social games in a socialnetwork.}
Phys. Rev. E {\bf 63}, 030901(R) (2001).



\bibitem{roc} C. P. Roca,  J. A.  Cuesta, A. S\'{a}nchez
%{\em Effect of spatial structure on the evolution of cooperation.}
Phys. Rev E {\bf 80}, 046106(1-16) (2009).


\bibitem{oht} H. Ohtsuki, C. Hauert, E. Lieberman, M. A. Nowak,
%{\em A simple rule for the evolution of cooperation on graphs and social networks.}
Nature {\bf 441}, 502-505 (2006).

\bibitem{doe} M. Doebeli and C. Hauert,
%{\em Models of cooperation based on the Prisoner's Dilemma and the Snowdrift game.}
Ecol. Lett. {\bf 8}, 748-766 (2005).

\bibitem{hauert} C. Hauert  and G. Szab\'{o},
%{\em Game Theory and Physics.}
Am. J. Phys. {\bf 73}, 405-414 (2005).

\bibitem{con} R. Cong, Y. Qiu, X. Chen, L.  Wang, 
%{\em Robustness of cooperation on highly clustered scalefree networks.}
Chin Phys Lett {\bf 27}, 030203/1-4, (2010).

\bibitem{ass} S. Assenza, J. G\'{o}mez-Garde\~{n}es, V. Latora,
%{\em Enhancement of cooperation in highly clustered scale-free networks}
Phys Rev E {\bf 78}, 017101 (2008).

\bibitem{lei} C. Lei, J. Jia, X. Chen, R. Cong, L. Wang,
%{\em Prisoner's dilemma game on clustered scale-free networks under different initial distributions.}
Chin Phys Lett {\bf 26}, 080202 (2009).

\bibitem{ift} M. Ifti, T. Killingback, M.  Doebeli,  J. Theor. Biol. {\bf 231}, 97-106 (2004).
%Effects of neighbourhood size and connectivity on prisoner"s dilemma

\bibitem{san2} F. C.  Santos, J. F.  Rodrigues, J.M.  Pacheco,
%Epidemic spreading and cooperation dynamics on homogeneous small-world networks.
Phys. Rev. E {\bf 72} 056128 (2005).

\bibitem{vuk2} J. Vukov, G. Szab\'o, A. Szolnoki, Phys. Rev. E {\bf 77}, 026109 (2008).
%Evolutionary Prisoner's Dilemma game on the Newman-Watts networks


\bibitem{gan} G. Wu, K. Gao, H. Yang, B.  Wang, Chin. Phys. Lett. {\bf 25}, 2307 (2008).
%Role of Clustering Coefficient on Cooperation Dynamics in Homogeneous Networks


\bibitem{was} S. Wasserman, S. and K. Faust, {\it Social Network Analysis:
Methods and Applications.} (1994, Cambridge University Press).

\end{thebibliography}
\end{document}